\documentclass[twocolumn,traditabstract]{aa}  
\usepackage{natbib}
\bibpunct{(}{)}{;}{a}{}{,} 
\usepackage{graphicx}
\usepackage{multirow}
\usepackage[colorlinks=true, allcolors=blue]{hyperref}
\usepackage[normalem]{ulem}
\usepackage{txfonts}
\begin{document} 

\title{CLIcK: a Continuum and Line fItting Kit for circumstellar disks} 

  \author{Yao Liu \inst{1,2}
          \and
          Ilaria Pascucci \inst{1,3,4}
          \and
          Thomas Henning \inst{1}
          }
  \institute{Max Planck Institute for Astronomy, K\"onigsthul 17, 69117, Heidelberg, Germany\\
              \email{yliu@mpia.de}
         \and
             Purple Mountain Observatory, Chinese Academy of Sciences, 2 West Beijing Road, Nanjing 210008, China
         \and
             Lunar and Planetary Laboratory, University of Arizona, Tucson, AZ 85721, USA 
         \and
             Earths in Other Solar Systems Team, NASA Nexus for Exoplanet System Science}

\authorrunning{Liu et al.}
  \titlerunning{Continuum and Line FItting Kit}

 
\abstract{Infrared spectroscopy with medium to high spectral resolution is essential to characterize the 
gas content of circumstellar disks. Unfortunately, conducting continuum and line radiative transfer of 
thermochemical disk models is too time-consuming to carry out large parameter studies. Simpler approaches 
using a slab model to fit continuum-subtracted spectra require the identification of either the global 
or local continuum. Continuum subtraction, particularly when covering a broad wavelength range, is 
challenging but critical in rich molecular spectra as hot (several hundreds K) molecular emission lines 
can also produce a pseudo continuum. In this work, we present CLIcK, a flexible tool to simultaneously fit the 
continuum {\it and} line emission. The DDN01 continuum model \citep{dullemond2001} and a plane-parallel 
slab of gas in local thermodynamic equilibrium are adopted to simulate the continuum and line emission respectively, 
both of them are fast enough for homogeneous studies of large disk samples. We applied CLIcK to fit the observed 
water spectrum of the AA\,Tau disk and obtained water vapor properties that are consistent with 
literature results. We also demonstrate that CLIcK properly retrieves the input parameters used to 
simulate the water spectrum of a circumstellar disk. CLIcK will be a versatile tool for the interpretation 
of future James Webb Space Telescope spectra.}   


\maketitle
%

\section{Introduction}
\label{sec:intro}

Circumstellar disks are a natural outcome of the star formation process \citep[e.g.,][]{shu1987} and 
provide the raw material to form planets \citep[e.g.,][]{lissauer2007}. Hence, clarifying how disks 
evolve and disperse is crucial to our understanding of planet formation. Although gas dominates the 
initial disk mass and dust accounts for only $\sim$1\%, most of our understanding of disk evolution 
and dispersal is based on the easier to detect dust component \citep[e.g.,][]{williams2011,ercolano2017}.

Infrared (IR) spectroscopy with the infrared spectrograph \citep[IRS,][$R\,{\sim}\,600$]{houck2004} on 
board the {\it Spitzer} Space Telescope opened a new window to investigate the gas content of disks. 
It revealed that the surface of disks around young solar analogues is rich in molecular emission, including 
water \citep[e.g.,][]{pontoppidan2014}, as well as atomic and ionic species \citep[e.g.,][]{lahuis2007,pascucci2007}. 
High resolution ($R\,{\sim}\,25,000\,{-}\,95,000$) ground-based follow-up of the very brightest disks 
demonstrated that these IR lines trace the terrestrial and giant planet forming region out 
to ${\sim}\,10$\,AU \citep[][]{pascucci2011,mandell2012}. Hence, they are crucial to map the evolution 
of volatiles \footnote{Molecules and atoms with low sublimation temperatures ($<$ a few 100K), including water.} while planets are assemblying.

Several interesting trends have emerged from the analysis of medium-resolution {\it Spitzer} spectra. 
Unlike young solar analogues, disks around very low-mass stars and sub-stellar objects have spectra with 
stronger emission from C-bearing molecules than water \citep{pascucci2009,pascucci2013} while disks around 
more massive Herbig Ae/Be stars lack detectable molecular emission \citep{pontoppidan2010} \footnote{The only 
exception is the CO$_2$ detection toward HD\,101412.}. In addition, the water line fluxes between 
$2.9\,{-}\,33$\,$\mu$m correlate with the size of the inner disk gap traced by CO ro-vibrational lines 
suggesting that the infrared water spectrum can be used to probe inside-out water depletion in the 
planet-forming zone \citep{banzatti2017}. Because line strengths depend not only on abundance but also 
on excitation, modeling is necessary to retrieve physical parameters such as column densities.

A few groups have carried out in depth physical and/or chemical models of a handful of well-known disks 
to match the dust as well as the atomic and molecular emission \citep[e.g.,][]{gorti2011,kamp2013,blevins2016}. 
The most recent work in this area  uses a new efficient and fast line raytracer \texttt{FLiTs} to post-process 
the thermochemical models of \texttt{ProDiMo} \citep{woitke2018}. Such studies, although extremely valuable 
to constrain the properties of specific disks, are too time-consuming to investigate typical disk properties 
in a star-forming region and explore how those properties evolve with time. As such, other groups have taken 
the much simpler approach of: first, subtracting the broad continuum emission; and then, finding the 
best-fit (with the minimum $\chi^2$) synthetic spectrum assuming a simple plane-parallel slab model of 
gas with a single temperature and column density \citep[e.g.,][]{carr2008,banzatti2012}. Interestingly, even 
adopting this simpler approach, molecular abundances reported in the literature differ from a few up to an 
order of magnitude, see e.g., the water column density inferred for the AA\,Tau disk from \citet{carr2011} 
and \citet{salyk2011}. These differences result from slightly different continuum subtractions, model 
assumptions (e.g., local line width), and choice of lines to fit. The continuum subtraction is especially 
critical in rich molecular spectra as hot (several hundreds K) molecular emission lines can also produce a 
pseudo continuum that cannot be accounted for with current approaches.

In this work, we develop a different framework to simultaneously fit the continuum {\it and} molecular 
emission spectra, thus accounting for any pseudo continuum produced by hot molecular lines. Our Continuum 
and Line fItting Kit (CLIcK) is presented in Sect.~\ref{sec:framework}. In Sect. \ref{sec:aatau}, we use 
our tool to fit the rich {\it Spitzer} water spectrum of the AA~Tau disk. We also check the quality of the 
fit and compare the derived gas properties with literature values that are obtained from different fitting 
approaches. In Sect. \ref{sec:refmodel}, we derive synthetic spectra of a known reference model, and fit 
the water lines to further demonstrate the applicability of our kit. The paper ends with a brief summary 
and outlook in the James Webb Space Telescope (JWST) era (Sect. \ref{sec:summary}).

\section{Framework of our Continuum and Line fItting Kit}
\label{sec:framework}

There are already several workflows to simulate dust continuum and line emission in the IR regime. 
For instance, \citet{Pontoppidan2009} developed a line raytracer \texttt{RADLite} that is well coupled 
with the continuum radiative transfer package \texttt{RADMC} \citep{dullemond2004a}. Dust continuum can 
be obtained along with the process of line raytracing. Since gas heating and cooling balance is not 
considered and there is no chemical network, both the gas temperature and abundance distribution of 
molecules must be provided either via parametric forms or from sophisticated models. Generally, one to 
two hours are needed to simulate a full spectrum of ${\sim}\,1,000$ water lines in the IR wavelength 
domain \citep{Pontoppidan2009}. 

Recently, \citet{woitke2018} ran full two-dimensional \texttt{ProDiMo} thermochemical disk models to 
calculate the dust and gas temperature structures, dust continuum and molecular concentrations. They 
post-processed the \texttt{ProDiMo} outputs with the fast line raytracer \texttt{FLiTs} \footnote{http://www.michielmin.nl/codes/flits/} 
to simulate the molecular line spectra in the IR. A full treatment of the chemical network requires global 
iterations of the disk structure, which is very time-consuming even running parallel on clusters \citep{woitke2016}.    

The huge computational cost in both workflows does not favor exploring a large range of parameters when 
fitting spectra. Therefore, to simulate molecular line emission, we adopt a simpler approach of a slab 
model in local thermodynamic equilibrium (LTE) as widely adopted in previous analysis of {\it Spitzer} 
spectra of circumstellar disks \citep[e.g.,][]{carr2008,salyk2011,banzatti2012,pascucci2013}. 
Particularly, our slab model is built upon the one developed by \citet{pascucci2013}, with molecular 
parameters updated to the 2016 edition of the HITRAN database \citep{gordon2017}. The slab of gas is 
characterized by the temperature $T$, the molecular line-of-sight column density $N$, and the projected 
emitting area $A$. We define $R_{e}$ as the radius of a circular region with this emitting area. 
Given the parameters \{$T$, $N$, $R_{e}$\}, a synthetic spectrum can be computed as detailed in the 
Appendix of \citet{banzatti2012}. The local line width is assumed to be only due to thermal broadening. 
In addition to an isothermal (single temperature) slab, we also introduce a range of temperatures 
through a power-law temperature profile. 

As mentioned above, continuum radiative transfer simulations are commonly carried out to reproduce 
spectral energy distributions (SEDs). The computation time depends strongly on the optical depth: for a typical 
circumstellar disk it is about 20 minutes without solving for vertical  hydrostatic equilibrium \citep{dullemond2004a}. 
This time is too long to generate a large number of models, which are necessary to fit the observed 
continuum especially when covering a broad wavelength range. \citet{chiang1997} (hereafter CG97) derived 
hydrostatic, radiative equilibrium models for passive circumstellar disks. CG97 models are characterized 
by an optically thin surface layer of superheated dust grains and a cool interior region. While the surface 
layer contributes most to the IR flux, the millimeter emission is dominated by the disk interior. 
Dullemond, Dominik and Natta  \citep[][hereafter DDN01]{dullemond2001} modified the CG97 model by 
including a puffed-up inner rim that produces a conspicuous bump at $2\,{-}\,3\,\mu{\rm m}$ in the 
SED. This rim emission was proposed to explain the near-IR bump observed 
in the SEDs of Herbig Ae stars. 

\begin{figure}[!t]
  \centering
  \includegraphics[width=0.47\textwidth]{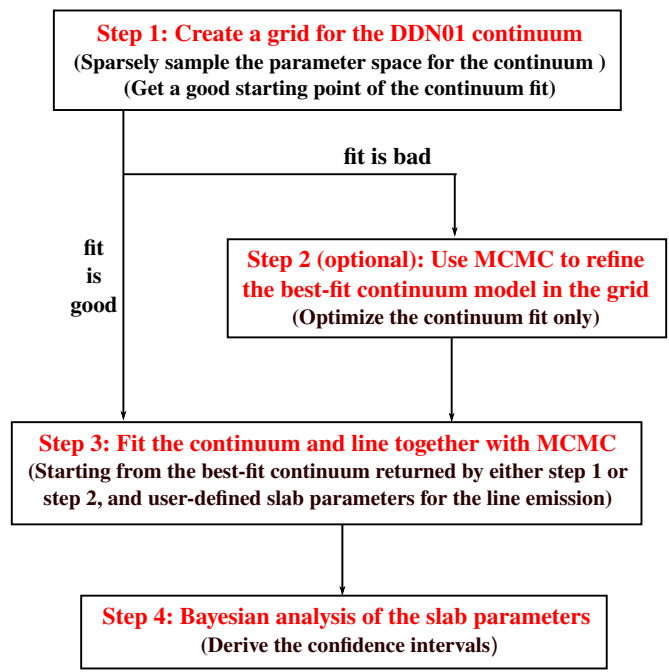}
  \caption{The flowchart of our fitting tool CLIcK.}
  \label{fig:flowchart}
\end{figure}

CLIcK calculates the dust continuum based on the DDN01 model which completes in just a few seconds. 
Details about the numerical solution are presented in \citet{dullemond2001}. In the fitting procedure, 
we explore the following parameters for the continuum: stellar luminosity, disk inner radius, the 
vertical height of the inner rim, the power-law index of the dust surface density profile, the total 
dust mass, and the disk inclination. The dust grains are composed of amorphous silicate and carbon, 
with complex refractive indices given by \citet{dorschner1995} and \citet{jager1998}. Mie theory is 
used to calculate the dust opacities. Different abundances of silicate and carbon and grain sizes are 
also considered, because they have a significant impact on the shape of the dust emission features 
such as the $10\,\mu{\rm m}$ silicate feature. The final model spectrum is a sum of the continuum 
and line emission. We neglect absorption of continuum radiation by molecules. Taking water as
an example, a vast majority of mid-IR water lines would be optically thin if the column density of the slab 
is around $10^{18}\,\rm{cm^{-2}}$, i.e., a typical value found for circumstellar disks \citep[e.g.,][]{carr2008,salyk2011}.
Therefore, the extinction induced by molecules will not significantly change the overall shape and level 
of the continuum. As shown in detailed 2D models \citep[e.g.,][]{bruderer2015,bosman2017}, most of the line 
fluxes come from a region where the dust optical depth $\tau_{\rm dust}\,{<}\,1$. Therefore, absorption of 
line emission by the dust is also ignored. Absorption of line radiation by the gas within the slab is taken 
into account in the same way as detailed in the Appendix of \citet{banzatti2012}. Under these assumptions, 
CLIcK is flexible enough to simultaneously include multiple molecules with different sets of slab 
parameters into the fitting process.

\begin{figure}[!t]
  \centering
  \includegraphics[width=0.48\textwidth]{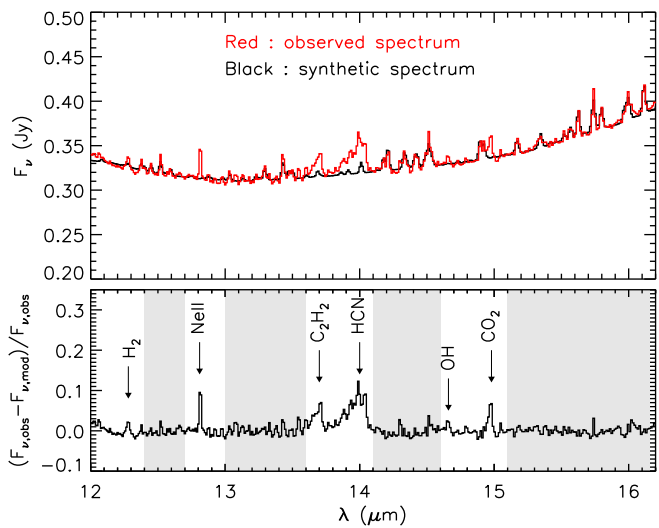}
  \caption{A comparison between the observed (red) and best-fit synthetic (black) spectrum in the 
  wavelength range $12\,{-}\,16\,\mu{\rm m}$. The relative difference between model and 
  observation $(F_{\nu,\rm{obs}}-F_{\nu,\rm{mod}})/F_{\nu,\rm{obs}}$ is shown in the bottom panel. 
  The gray shaded regions mark the data points that contribute to the $\chi^2$ calculation in fitting 
  the water spectrum. Prominent emission lines from other molecules  are indicated.} 
  \label{fig:bestfitsh}
\end{figure}

The main steps of our fitting procedure, as implemented in CLIcK, are summarized in Figure~\ref{fig:flowchart}. 
Since the overall flux level of a spectrum is generally characterized by the continuum, the first task (Step~1) 
of CLIcK is to search for a good continuum model as the starting point. To do this, we create a coarse DDN01 
model grid and identify the best fit based on the $\chi^2$ metric. A sparse mapping of the parameter space not 
always provides an acceptable fit. Therefore, the second module (Step~2) invokes the Markov Chain Monte Carlo (MCMC) 
algorithm to improve upon the best fit returned by the grid. Details about the MCMC implementation can be found 
in the Appendix of \citet{liu2012} and \citet{liu2013}. Starting from the continuum model and user-chosen 
input parameters for the slab of gas, CLIcK optimizes the continuum {\it and} line together (Step~3) again 
adopting the MCMC approach \footnote{This approach is advantageous in high dimensional modeling such as those 
explored here.}. As the last step (Step~4), we conduct a Bayesian analysis of the slab parameters to derive 
their confidence intervals. CLIcK is intentionally devised to efficiently work on a personal computer that 
has just a few cores. The application of our hybrid fitting framework is a compromise between the fitting 
quality, required time, and limited computational resources \citep{liu2013}.

\section{CLIcK on the observed spectrum of AA~Tau}
\label{sec:aatau}

AA\,Tau is a classical T Tauri star located in the Taurus star formation region at a distance 
of ${\sim}137\,\rm{pc}$ \citep{gaia2016,gaia2018}. The spectral type of the star is M0.6 and the stellar 
mass ${\sim}\,0.6\,M_{\odot}$ \citep{herczeg2014}. AA\,Tau has been the focus of intense multi-wavelength 
campaigns due its periodic dimming which is ascribed to an extended non-axisymmetric feature, such as a 
disk warp, passing in front of a star+disk system seen close to edge-on \citep[e.g.,][]{bouvier2013}. 
Recent ALMA continuum images at ${\sim}\,0.2^{\prime\prime}$ resolution have revealed that the outer disk 
is rather inclined with respect to the observer ($59^{\circ}$) but not edge-on \citep{loomis2017}, providing 
evidence for an inner vs. outer disk misalignment as responsible for the most recent long-duration dimming 
that started in 2011 \citep[e.g.,][]{rodriguez2015}. Before this dimming, AA~Tau was observed with the 
{\it Spitzer}/IRS at a spectral resolution of $R\,{\sim}\,600$. \citet{carr2008} were first to report 
its molecular spectrum from 10 out to $37\,\mu{\rm m}$ which is particularly rich in water emission lines. 
As such, AA\,Tau is an excellent target to test and showcase our continuum and line fitting kit. 

\begin{figure}[!t]
  \centering
  \includegraphics[width=0.49\textwidth]{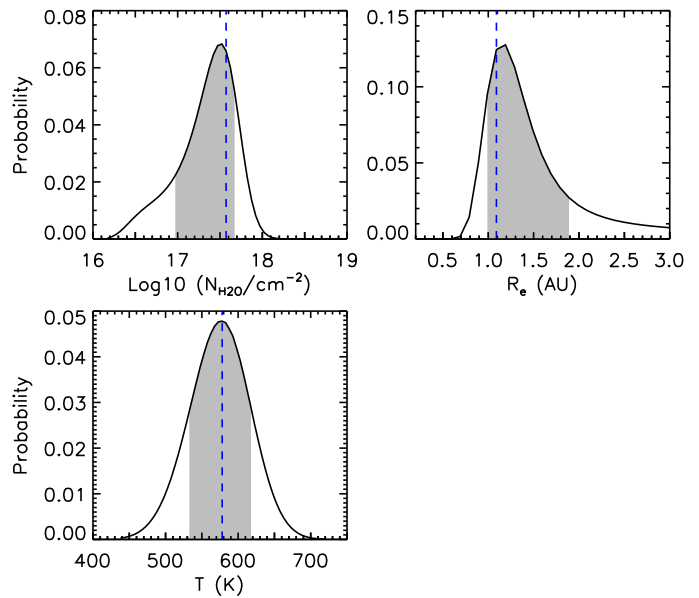}
  \caption{Bayesian probability distributions of the ${\rm H_2O}$ parameters for AA\,Tau. These distributions 
  are used to evaluate the 68\% confidence intervals (shaded regions) of the best-fit model presented in 
  Figure~\ref{fig:bestfitsh}. The blue vertical dashed lines give the best-fit values.}
  \label{fig:shfitpdf}
\end{figure}

\begin{table*}[!t]
\caption{Modeling results obtained when fitting the $12\,{-}\,16\,\mu{\rm m}$ $\rm{H_2O}$ lines in the AA\,Tau IRS spectrum.}
\centering
\linespread{1.3}\selectfont
\begin{tabular}{lcccc}
\hline 
\hline
Reference        & $N_{\rm H_{2}O}\,(\rm{cm}^{-2})$   &  T\,(K)     &   $R_{e}\,\rm{(AU)}$  & $\chi^2$ \\
\hline 
\citet{carr2008} & $6.5\pm2.4\times10^{17}$      &  $575\pm50$ &   $2.1\pm0.1$  & --- \\ 
\hline
\citet{carr2011} & $7.8\pm2\times10^{17}$        &  $575\pm50$ &   $0.85\pm0.12$ & --- \\   
\hline
This work        & $3.7^{+1.0}_{-2.5}\times10^{17}$   &  $577^{+35}_{-45}$   &   $1.1^{+0.6}_{-0.1}$ & 64 \\
\hline
\end{tabular}
\linespread{1.0}\selectfont
\tablefoot{(a) $R_{e}$ is the radius of emitting area. (b) The slab is assumed to be isothermal in all of these 
three studies. (c) The errors given along with our best-fit values corresponds to the $1\,\sigma$ confidence interval.}
\label{tab:shfit}
\end{table*}

\begin{table*}[!t]
\caption{Modeling results obtained when fitting the $19\,{-}\,32\,\mu{\rm m}$ $\rm{H_2O}$ lines in the AA\,Tau IRS spectrum.}
\centering
\linespread{1.3}\selectfont
\begin{tabular}{lcccc}
\hline 
\hline
Reference        & $N_{\rm H_{2}O}\,(\rm{cm}^{-2})$   &  T\,(K)     &   $R_{e}\,\rm{(AU)}$ & $\chi^2$ \\
\hline
This work        & $4.1^{+3.21}_{-0.85}\times10^{17}$   &  $397^{+15}_{-25}$   &   $1.5^{+0.1}_{-0.1}$ & 500 \\
\hline
\end{tabular}
\label{tab:lhfit}
\end{table*}

\begin{table*}[!t]
\caption{Modeling results obtained when fitting almost the entire IRS water spectrum ($12\,{-}\,32\,\mu{\rm m}$) of AA\,Tau.}
\centering
\linespread{1.3}\selectfont
\begin{tabular}{lccccc}
\hline 
\hline
Reference    &  $N_{\rm H_{2}O}\,(\rm{cm}^{-2})$   &  $T$\,(K)   &   $R_{e}\,\rm{(AU)}$  & $P$ & $\chi^2$  \\
\hline
\citet{salyk2011} & $6.3\times10^{18}$   &  400   &   0.74  & 0 &  ---   \\
\hline
This work     & \multirow{2}{*}{$1.0^{+0.2}_{-0.2}\times10^{18}$} & \multirow{2}{*}{$442^{+15}_{-15}$} & \multirow{2}{*}{$1.23^{+0.1}_{-0.1}$} & \multirow{2}{*}{0} & \multirow{2}{*}{745} \\
(isothermal slab)               \\ 
\hline
This work     & \multirow{2}{*}{$1.27^{+0.15}_{-0.37}\times10^{18}$} & \multirow{2}{*}{$368^{+10}_{-5}$} (a) & \multirow{2}{*}{$1.79^{+0.1}_{-0.1}$} & \multirow{2}{*}{$-0.43^{+0.05}_{-0.05}$} & \multirow{2}{*}{700} \\
(Power-law profile for $T$) &         \\
\hline
\end{tabular}
\label{tab:fulfit}
\tablefoot{(a): This value refers to the characteristic temperature at 1\,AU ($T_{1\,\rm{AU}}$), given the 
power-law profile of $T\,{=}\,T_{1\,\rm{AU}}\,{\times}\,R^{p}$.}
\end{table*}

\subsection{Fit to the short-wavelength IRS spectrum}\label{sec:fitsh}
The short-wavelength IRS module covers from $\sim$10 out to 20\,$\mu{\rm m}$. \citet{carr2008} and \citet{carr2011} 
modeled the water emission spectrum of AA\,Tau in the wavelength range $12\,{-}\,16\,\mu{\rm m}$ using a slab of 
gas in LTE. They first subtracted the observed spectrum with a user-defined continuum and then performed the 
line fitting. Our approach differs from \citet{carr2008} and \citet{carr2011} in that we fit the continuum 
and line emission together.

First, we ran a grid of dust continuum SED models to identify the one that roughly matches the observed flux 
density versus wavelength. Then, starting from the best continuum model, together with a set of initial parameters 
for the $\rm{H_2O}$ slab, we find the best fit (with the minimum $\chi^2$) to the continuum {\it and} water lines 
by calculating 8 MCMCs, each of which containing 2,000 models. The parameters determining both the continuum and 
water lines are allowed to vary during the optimization. Since we are more interested to interpret the line 
emission, data points at wavelengths dominated by emission lines from other species (for instance HCN, 
$\rm{C_2H_2}$, $\rm{CO_2}$ and OH) or by continuum alone do not contribute to the resulting $\chi^2$. Our 
best-fit model to the $12\,{-}\,16\,\mu{\rm m}$ spectrum of AA\,Tau is shown in black in the upper panel of 
Figure~\ref{fig:bestfitsh}. The lower panel of Figure~\ref{fig:bestfitsh} presents the difference between 
the best fit model and the observed spectrum, with the gray shaded regions indicating the wavelengths that 
contribute to the $\chi^2$ calculation. Note how the model reproduces well the observation, with 
relative differences typically less than 5\% for the full range of fitted wavelengths.

To determine the confidence intervals of the best-fit parameters, we ran a dense grid of models with the gas 
temperature varying between $T_{\rm best}\,{-}\,150\,\rm{K}$ and $T_{\rm best}\,{+}\,150\,\rm{K}$ (in steps of 5\,K), 
column density between ${\rm Log}(N_{\rm best})\,{-}\,1.5$ and ${\rm Log}(N_{\rm best})\,{+}\,1.5$ (in logarithmic 
steps of 0.05) and emitting area between $R_{\rm e,best}\,{-}\,1.5\,\rm{AU}$ and $R_{\rm e,best}\,{+}\,1.5\,\rm{AU}$ 
(in steps of 0.1\,AU), where the subscript ``best'' refers to the best-fit values. Then, a Bayesian analysis was 
performed to calculate the probability distribution for each parameter (Figure~\ref{fig:shfitpdf}), from which we 
compute 68\% of the area under the functions to assign 1$\sigma$ uncertainties. We place strong constraints on the 
water column density and temperature, as illustrated by the clear peaks in the corresponding probability distributions 
and by the minor discrepancies between these peaks and the best-fit values (see Figure~\ref{fig:shfitpdf}). 
Within 2$\sigma$, our best-fit parameters are the same as those reported in \citet{carr2008} and \citet{carr2011} 
(see Table~\ref{tab:shfit}), demonstrating the applicability of our fitting kit. 

\begin{figure}[!t]
  \centering
  \includegraphics[width=0.48\textwidth]{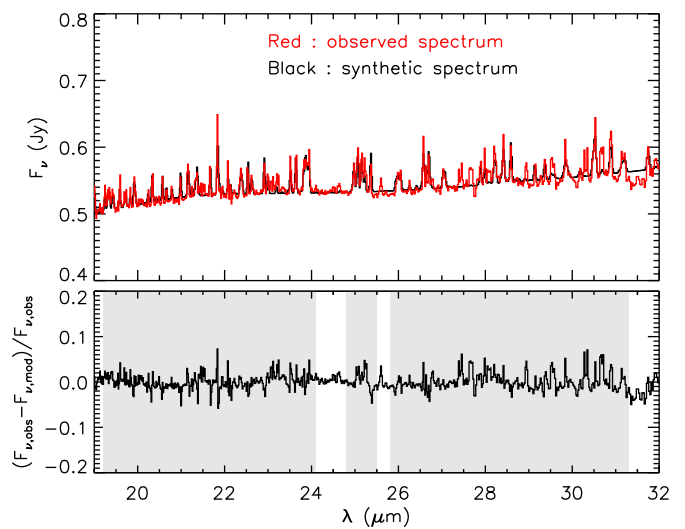}
  \caption{Same as in Figure~\ref{fig:bestfitsh}, but for a fit to the water lines in the wavelength range from 
  19 to $32\,\mu{\rm m}$.} 
  \label{fig:bestfitlh}
\end{figure}

\subsection{Fit to the long-wavelength IRS spectrum}

Using the same optimization process described above, we fit the water lines shown in the IRS spectrum from 19 
to $32\,\mu{\rm m}$. The best-fit synthetic spectrum is in good agreement with the observed one, see Figure~\ref{fig:bestfitlh}. 
We find that the best-fit temperature ($T\,{=}\,397^{+15}_{-25}\,\rm{K}$, Table~\ref{tab:lhfit}) is lower than the 
one ($T\,{=}\,577^{+35}_{-45}\,\rm{K}$, see Table~\ref{tab:shfit}) obtained from fitting the spectrum at shorter 
wavelengths, while the emitting radius is slightly larger, though the same within the uncertainty. 
This is as expected because emission lines at shorter wavelengths predominately trace a hotter disk region that 
is closer to the central star.  

\subsection{Fit to the entire IRS spectrum}

Finally, we tested our tool by fitting almost the entire water spectrum, from 12 to $32\,\mu{\rm m}$. This has been 
challenging for previous methods (see Sect. \ref{sec:intro}) as it is difficult to correctly identify the global 
continuum in rich molecular spectra obtained at relatively poor spectral resolution. Indeed, the only attempt to 
retrieve molecular parameters for the entire IRS spectrum was made by \citet{salyk2011}. Instead of identifying 
the global continuum and fitting the entire spectrum, \citet{salyk2011} identified 65 relatively isolated water 
lines, subtracted local continua, and fit their peak emission assuming a slab of gas in LTE with one temperature. 
Here, we use our tool to fit the continuum {\it and} water lines adopting two temperature profiles: a constant 
temperature (isothermal slab) and a radially decreasing power-law profile.
 
\begin{figure}[!t]
  \centering
  \includegraphics[width=0.48\textwidth]{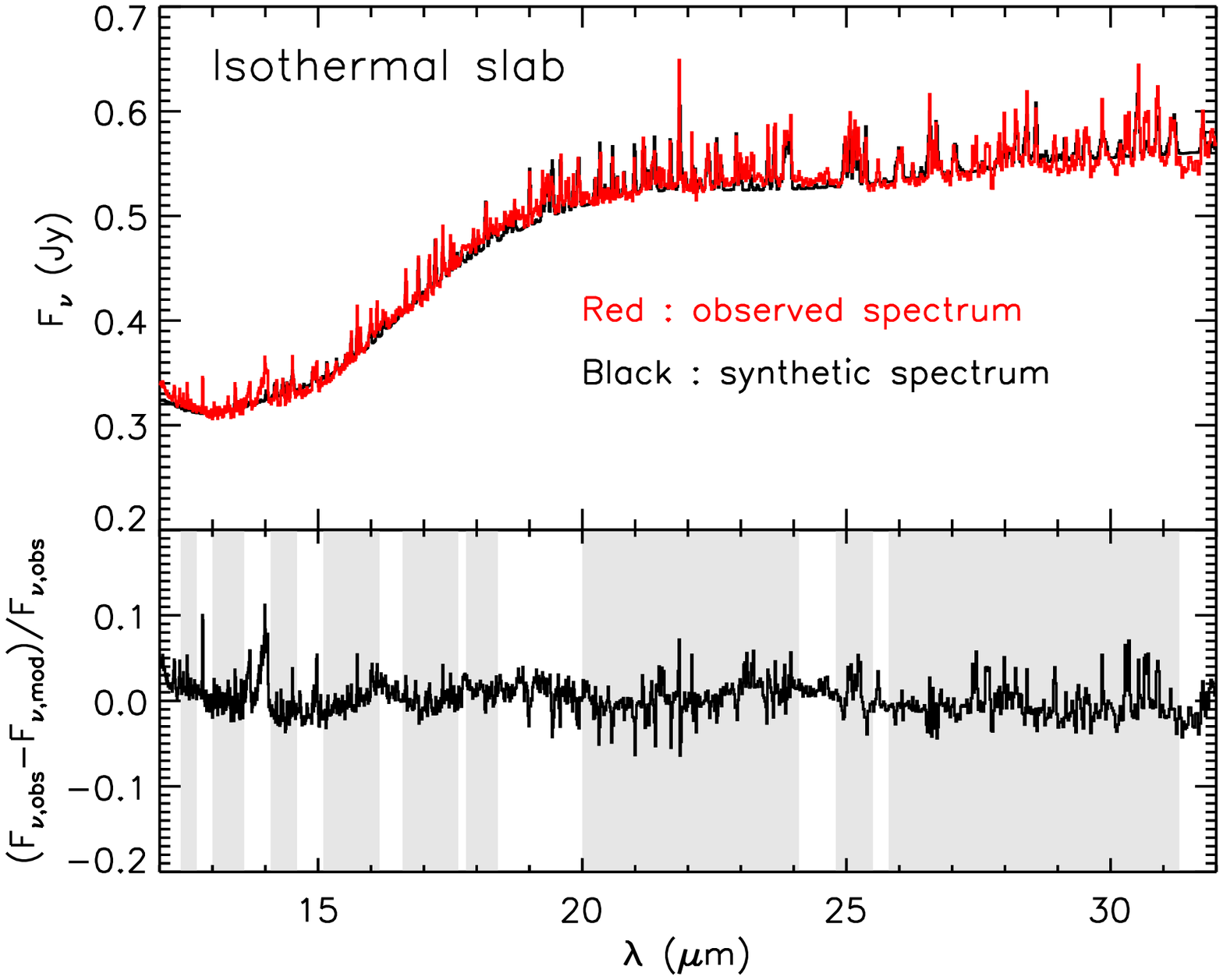}
  \includegraphics[width=0.48\textwidth]{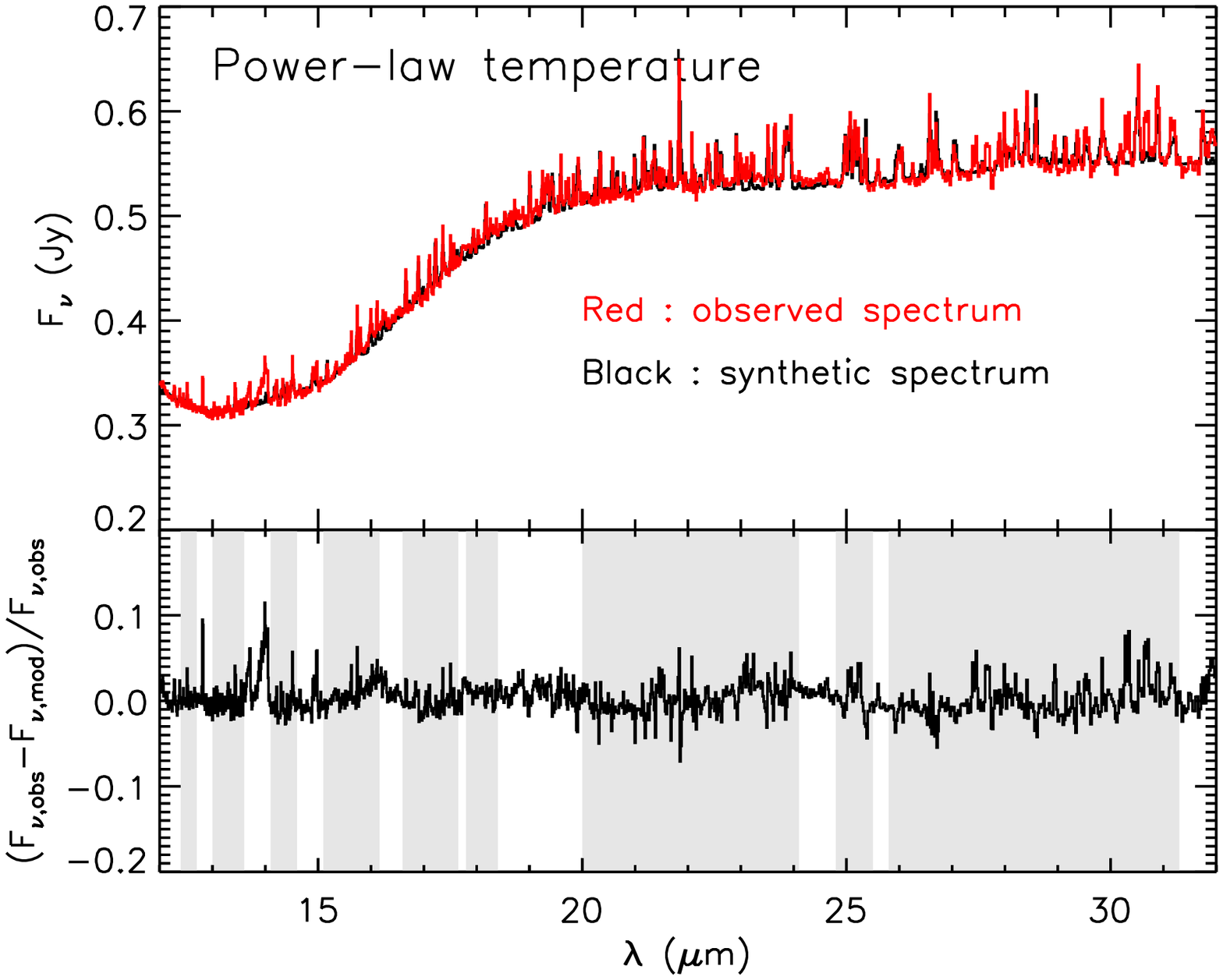}
  \caption{Same as in Figure~\ref{fig:bestfitsh}, but the IRS spectrum of a broad wavelength range from 12 to 
  $32\,\mu{\rm m}$ is considered in the fitting procedure. The temperature in the slab is assumed to be constant 
  (upper panel) and follow a radially decreasing power-law profile $T\,{=}\,T_{1\,\rm{AU}}\,{\times}\,R^{p}$ 
  (bottom panel), respectively.}
  \label{fig:bestful}
\end{figure}

\begin{figure}[!t]
  \centering
  \includegraphics[width=0.48\textwidth]{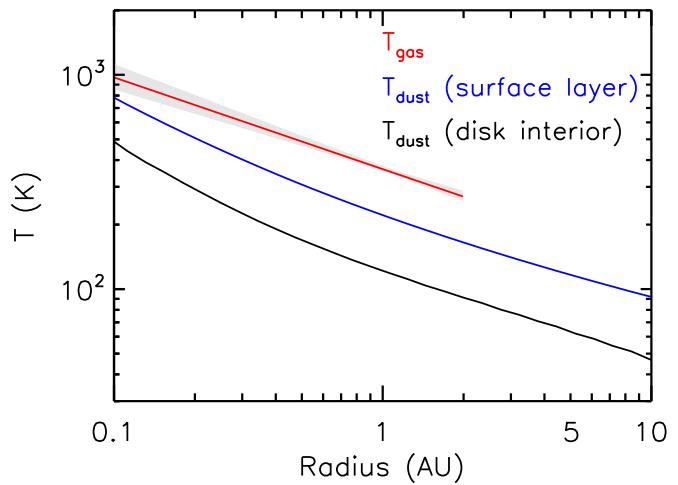}
  \caption{The temperature distribution for the gas (red), dust in the surface layer (blue) and disk interior (black) 
  of the model presented in the bottom panel of Figure~\ref{fig:bestful}. The shaded region associated with the gas 
  temperature indicates the confidence interval, see Table~\ref{tab:fulfit}.}
  \label{fig:bestful-temp}
\end{figure}

The upper panel of Figure~\ref{fig:bestful} shows a comparison between the best-fit synthetic spectrum for the isothermal 
slab and the observation while Table~\ref{tab:fulfit} reports the best-fit parameters. Our tool does a good job in reproducing 
the continuum and line emission as the relative difference between the synthetic and observed spectrum is smaller than 5\% for most 
of the considered wavelengths. The isothermal slab best-fit temperature ($T\,{=}\,442^{+15}_{-15}\,\rm{K}$) and 
emitting area ($R_{e}\,{=}\,1.23^{+0.1}_{-0.1}\,\rm{AU}$) are intermediate between those obtained by fitting only 
the short-wavelength (see Table~\ref{tab:shfit}) and only the long-wavelength (see Table~\ref{tab:lhfit}) portions 
of the spectrum. This is as expected and further demonstrates that our tool is working properly. We also note that 
our best-fit temperature and emitting region are larger than those reported by \citet{salyk2011} while the water 
column density is lower, see Table~\ref{tab:fulfit}. \citet{salyk2011} commented that fitting continuum-subtracted 
spectra, instead of individual line peaks, results in higher temperature, which agrees with our finding, but also 
higher column density and smaller emitting area, which is the opposite of what we find. However, as the latter two 
quantities are degenerate \footnote{These two parameters can be considered as scaling factors for the line strength 
as long as the emission lines are optically thin.}, increasing column density coupled with a decrease in emitting 
area, or vice versa, will produce model fits of similar quality. 

The bottom panel of Figure~\ref{fig:bestful} show the fitting results for a radially decreasing power-law temperature 
profile. Observations with a broad wavelength coverage are sensitive to a range of gas temperatures. Consequently, this 
model provides an even better fit to the data than the isothermal slab as shown by the lower $\chi^2$, see Table~\ref{tab:fulfit}. 
Even if we take into account the fact that this new model has one more free parameter, the conclusion holds because 
the number of fitted data points is much larger than the dimensionality of the model. Figure~\ref{fig:bestful-temp} 
presents the gas and dust temperature of the best-fit model, both of which share a similar power-law exponent. 
Moreover, the gas is hotter than the dust in the surface layer, which is consistent with circumstellar disk models 
that properly treat heating and cooling \citep[e.g.,][]{najita2011,akimkin2013,du2014}.

\section{CLIcK on a reference model}
\label{sec:refmodel}

We further test our kit by generating and then fitting a model spectrum for a circumstellar disk located in the 
Taurus star-forming region. The properties of the reference model are characterized by a set of parameters which 
we set as a prior in the simulation. We will then discuss how well our tool reproduces the spectrum and retrieves 
the properties of the water emitting region. 

The reference model consists of a star surrounded by a disk populated with dust and gas. Details about the dust 
density structure, dust properties, and stellar parameters are given in Sect.~\ref{sec:contrt}. We first conduct 
continuum radiative transfer to self-consistently calculate the temperature structure of the dust using 
the \texttt{RADMC} package \citep{dullemond2004a}. Previous observations and modeling suggest that dust and 
gas are thermally decoupled particularly in disk surface layers \citep{kamp2004,akimkin2013,blevins2016}. 
Therefore, we scale the gas temperature relative to the dust temperature using the results from the thermo-chemical 
model of \citet{najita2011}, see Sect.~\ref{sec:linert} for details. The resulting gas temperature, together 
with an abundance distribution of relevant molecular species, form the input for the line radiative transfer 
that is performed with the \texttt{RADLite} code \citep{Pontoppidan2009}. Following previous tests, we will 
only consider water lines in the simulation, see Sect.~\ref{sec:linert} for further details on the water 
abundance distribution and line radiative transfer. Finally, the output spectrum from \texttt{RADLite} is 
convolved with a spectral resolution of 2,000, the same offered by the Mid-Infrared Instrument (MIRI) on 
board JWST. 
 
\begin{figure}[!t]
  \centering
  \includegraphics[width=0.48\textwidth]{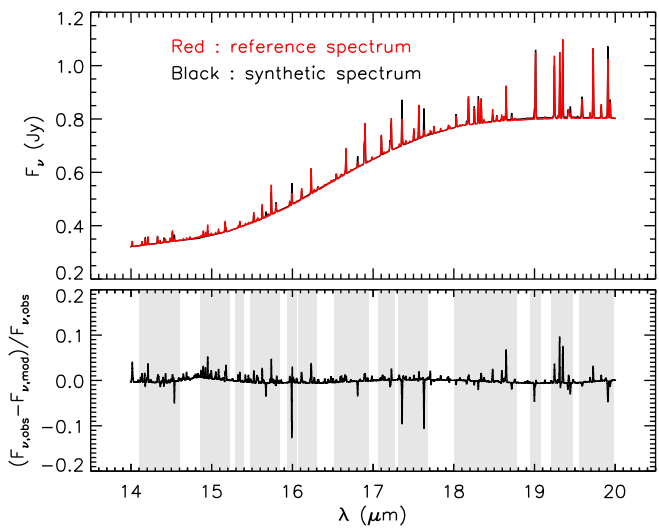}
  \caption{Comparison between the reference spectrum (red) and the best-fit synthetic spectrum (black) obtained 
  with CLIcK.}
  \label{fig:fitrefmod}
\end{figure}

\begin{figure}[!t]
  \centering
  \includegraphics[width=0.49\textwidth]{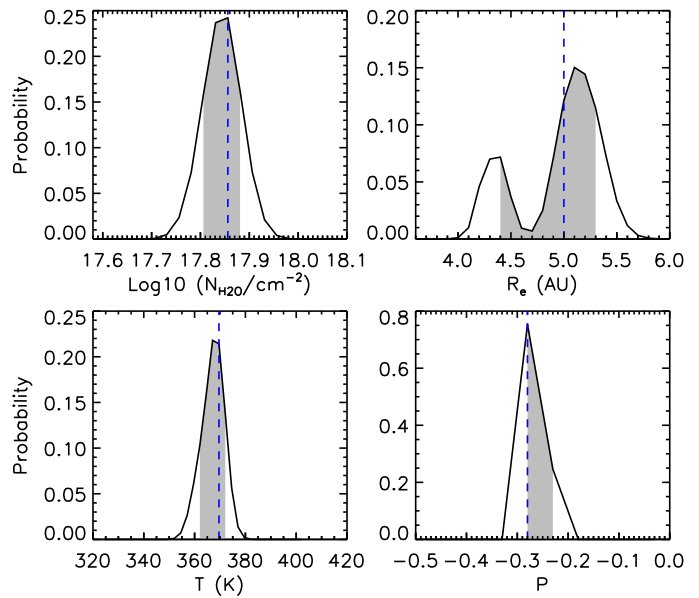}
  \caption{Bayesian probability distributions of the ${\rm H_2O}$ parameters for the fit to the reference spectrum. 
  Note that the abscissa scale in the $N_{\rm H_2O}$ and $T$ panels is different here from that in Figure~\ref{fig:shfitpdf}. 
  For the Bayesian analysis, we varied $T$ in steps of 2.5\,K, $N_{\rm H_2O}$ in logarithmic steps of 0.025, $R_{\rm e}$ in 
  steps of 0.1\,AU, and $P$ in steps of 0.05. The step widths for $T$ and $N_{\rm H_2O}$ are smaller than those used to 
  create Figure~\ref{fig:shfitpdf} (see Sect.~\ref{sec:fitsh}). These plots show that the strong peaks are not due to an 
  insufficient sampling of the parameters.}
  \label{fig:fitrefmodpdf}
\end{figure}

\begin{figure}[!t]
  \centering
  \includegraphics[width=0.48\textwidth]{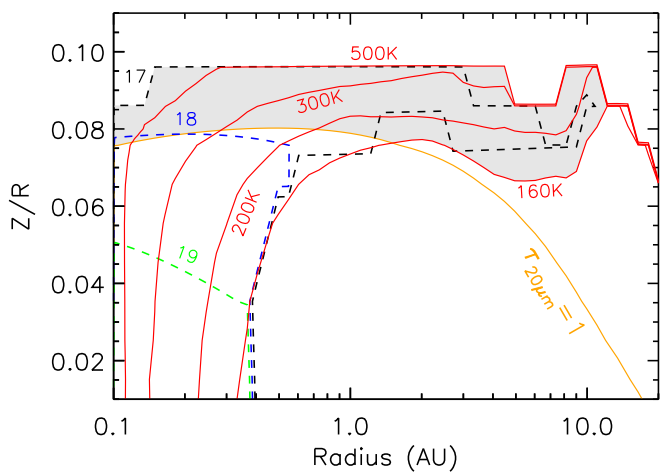}
  \caption{Gas temperature and $\rm{H_{2}O}$ column density distribution for the reference model. Contours of the gas temperature 
at 160, 200, 300 and 500\,K are shown with red solid lines. The dashed contours refer to the water column density ${\rm Log}\,(N_{\rm H_2O}/{\rm cm}^{-2})$ 
at 17 (black), 18 (blue) and 19 (green), where $N_{\rm H_2O}$ is measured from the surface layer down to the midplane. The orange contour indicates the scale 
height with dust optical depth $\tau_{20\,\mu\rm{m}}\,{=}\,1$ measured from the surface layer to the midplane. The shaded area encloses the contours 
of ${\rm Log}\,(N_{\rm H_2O}/{\rm cm}^{-2})\,{=}\,17$, $T\,{=}\,160,\,500\,\rm{K}$ and $\tau_{20\,\mu{\rm m}}\,{=}\,1$. This region contributes 
to most of the water vapor emission at mid-IR wavelengths.}
  \label{fig:gastemp}
\end{figure}

Using the same procedure described in Sect.~\ref{sec:aatau}, we fit the water lines of the reference model, with the results 
presented in Figures~\ref{fig:fitrefmod} and \ref{fig:fitrefmodpdf} and the best-fit parameters summarized in Table~\ref{tab:fitrefmod}. 
The black line in the upper panel of Figure~\ref{fig:fitrefmod} shows the best fit synthetic spectrum which indeed reproduces most of 
the water lines with relative differences less than 5\%. Figure~\ref{fig:gastemp} shows the gas temperature and water vapor column density 
of the reference model. On that figure we also draw the surface for dust optical depth at $20\,\mu{\rm m}$ equal to one ($\tau_{20\,\mu{\rm m}}\,{=}\,1$). 
The shaded area encloses the contours of ${\rm Log}\,N_{\rm H_2O}=17$, $T=160,\,500\,\rm{K}$ and $\tau_{20\,\mu{\rm m}}\,{=}\,1$. 
Therefore, these regions are expected to contribute to most of the water vapor emission at mid-IR wavelengths. Our best-fit column 
density ($N_{\rm H_{2}O}\,{=}\,7.2\times10^{17}\,{\rm cm^{-2}}$), emitting area ($R_{e}\,{=}\,5\,\rm{AU}$), and 
temperature ($T\,{=}\,370\,\rm{K}$, $P\,{=}\,-0.28$) match well with the water vapor properties in the disk regions where most of the 
emission comes out, see Figure~\ref{fig:gastemp}. This agreement shows that our tool is capable of correctly retrieving the 
properties of the emitting gas, even though it cannot constrain the vertical distribution of the molecules. The water vapor mass 
in the emitting region is ${\sim}\,40\%$ of the total water vapor mass in the reference model, in agreement with the fact that 
IR observations only probe a portion of the disk, e.g., the disk surface layers. 

\begin{table}[!t]
\caption{Retrieved water line parameters for the reference model.}
\centering
\linespread{1.3}\selectfont
\begin{tabular}{cccc}
\hline 
\hline
$N_{\rm H2O}\,(\rm{cm}^{-2})$   &  $T_{1\,\rm{AU}}$\,(K)     &   $R_{e}\,\rm{(AU)}$      &   $P$  \\
\hline
$7.2^{+0.4}_{-0.8}\times10^{17}$   &  $370^{+2}_{-8}$   &  $5^{+0.3}_{-0.6}$ &  $-0.28^{+0.05}_{-0.00}$   \\
\hline
\end{tabular}
\label{tab:fitrefmod}
\end{table}

\section{Summary and outlook}
\label{sec:summary}

We have developed a flexible tool (called CLIcK) to simultaneously fit the continuum {\it and} line emission from circumstellar 
disks. The DDN01 continuum model and a slab model of gas in LTE are used to simulate the continuum and line emission 
respectively, both of them are fast enough to enable large parameter studies. We applied CLIcK to fit the observed water 
spectrum from the AA\,Tau disk and found water vapor properties that are consistent with literature values obtained by fitting 
continuum-subtracted spectra with a slab of gas in LTE. However, CLIcK has two advantages over these past approaches: it does 
not require to pre-define a global or local continuum and, in the fitting procedure, it accounts for any pseudo-continuum 
arising from hot molecular emission lines. We have also demonstrated that CLIcK can reproduce well the entire {\it Spitzer}/IRS 
water emission spectrum from AA\,Tau, especially with the power-law temperature profile option, and correctly retrieve the 
parameters of the water emitting region from a reference spectrum simulated with \texttt{RADLite}.
  
The framework of CLIcK can be easily generalized to solve similar problems in which the dust continuum has to be properly 
treated, including fitting the emission features from crystalline silicates \citep[e.g.,][]{juhasz2010} and polycyclic 
aromatic hydrocarbons \citep[e.g.,][]{seok2017} in circumstellar disks.

Given the different spectral analysis used in the literature, re-fitting {\it Spitzer}/IRS archival spectra of circumstellar 
disks with CLIcK will be very valuable to quantify the water content in a homogeneous way. JWST will bring a remarkable leap 
forward in infrared spectroscopy. The spectrometers on board JWST, e.g., MIRI \citep{rieke2015} with a spectral resolution 
of $R\,{\sim}\,2,000$, will not only separate many blended lines but also boost line-to-continuum ratios, resulting in a 
better detection of individual lines. CLIck will be a useful tool to efficiently fit the water spectrum, subtract it, and 
then search for additional molecules that might be present in the atmosphere of inner disks. This will help building the 
chemical inventory of the terrestrial planet forming region around young stars. Moreover, with two orders of magnitude 
increase in sensitivity, JWST will be able to significantly increase the sample of faint disks around low-mass stars and 
brown dwarfs that could be observed with {\it Spitzer}. A homogeneous analysis of these spectra with CLIcK will clarify 
any dependencies between gas properties and stellar mass down to the brown dwarf regime.

\begin{acknowledgements}
We thank the anonymous referee for very constructive comments that improved the manuscript. YL acknowledges supports 
by the Natural Science Foundation of China (Grant No. 11503087) and by the Natural Science Foundation of Jiangsu Province of 
China (Grant No. BK20181513) and . We thank John S. Carr for providing the reduced {\it Spitzer}/IRS spectrum of AA\,Tau and Klaus Pontoppidan 
for help with the \texttt{RADLite} code. We acknowledge Ewine F. van Dishoeck for insightful discussions. This material 
is based upon work supported by the National Aeronautics and Space Administration under Agreement No. NNX15AD94G for the 
program “Earths in Other Solar Systems”. The results reported herein benefited from collaborations and/or information 
exchange within NASA’s Nexus for Exoplanet System Science (NExSS) research coordination network sponsored by NASA’s Science 
Mission Directorate.
\end{acknowledgements}

\bibliographystyle{aa}
\bibliography{click}

\begin{appendix}
\section{Simulate the infrared spectrum of the reference model}
To simulate our reference spectrum we used a continuum and a line radiative transfer code. In this Appendix, we provide details 
about both codes.

\subsection{Continuum Radiative Transfer}
\label{sec:contrt}

{\it Density structure:} \hspace*{0.2cm} we employ the \texttt{RADMC} package to solve the problem of continuum radiative 
transfer assuming a two-dimensional flared disk model \citep{dullemond2004a}. The model includes two distinct grain populations 
that differ in terms of dust opacity and vertical extension in the disk, i.e., a small grain population and a large grain 
population. The disk extends from an inner radius of 0.1\,au to an outer radius of 200\,au. The total dust mass in the disk 
is defined as $M_{\rm dust,tot}$.

Small dust grains occupy a minor fraction of the dust mass, $(1-f)\,M_{\rm dust.tot}$, and are distributed vertically following 
a power-law profile
\begin{equation}
h = h_{c}\times\left(\frac{R}{R_{c}}\right)^\beta.
\label{eq:heightgas}
\end{equation}
The degree of disk flaring is quantified with $\beta$ and $h_{c}$ represents the scale height at a fixed characteristic 
radius of $R_{c}\,{=}\,100\,\rm{au}$. Conversely, most of the dust mass is composed of large dust grains, $f\,M_{\rm dust.tot}$, 
which are concentrated close to the midplane with a scale height $\Lambda\,h$. The parameter $\Lambda\,{<}\,1$, describing the 
degree of dust settling, is assumed to be constant over radius $R$. This is a simple approach to account for the effect of dust 
sedimentation in protoplanetary disks \citep[e.g.,][]{dullemond2004b,dalessio2006}. 

We model the surface density profile as a power law with an exponential taper
\begin{equation}
\Sigma \propto \left(\frac{R}{R_{c}}\right)^{-\gamma}{\rm exp}\left[-\left(\frac{R}{R_{c}}\right)^{2-\gamma}\right],
\label{eqn:sigma}
\end{equation}
where $\gamma$ is the gradient parameter. 

The two-dimensional density structure for each grain population can be parameterized as   
\begin{equation}
\rho_{\rm{small}}\propto\frac{(1-f)\Sigma}{h}\,\exp\left[-\frac{1}{2}\left(\frac{z}{h}\right)^2\right], \\
\label{eqn:dens}
\end{equation}
\begin{equation}
\rho_{\rm{large}}\propto\frac{f\Sigma}{\Lambda h}\,\exp\left[-\frac{1}{2}\left(\frac{z}{\Lambda h}\right)^2\right], \\
\label{eqn:dens}
\end{equation}
where $\rho_{\rm{small}}$ is the density for the small grain population and $\rho_{\rm{large}}$ is the density for 
the large grain population. Our parameterization of the density structure has been used to successfully model 
multi-wavelength observations of protoplanetary disks in the literature \citep[e.g.,][]{andrews2011,fang2017,fedele2018}. 
Table~\ref{tab:refmodpara} summarizes the input parameters and the values we chose for the reference model.

~~~~~~~~~~~~~~~~\\
{\it Dust properties:} \hspace*{0.2cm} the dust in the model is a mixture of 75\% amorphous silicate and 25\% carbon, with 
complex refractive indices given by \citet{dorschner1995} and \citet{jager1998}. Mie theory is used to calculate the dust 
opacities. The grain size distribution follows the standard power law ${\rm d}n(a)\propto{a^{-3.5}} {\rm d}a$ with a 
minimum grain size of $a_{\rm{min}}=0.01\,\mu{\rm m}$. The maximum grain size is set to $a_{\rm{max}}\,{=}1\,\mu\rm{m}$ 
for the small grain population and $a_{\rm{max}}\,{=}\,1\,\rm{mm}$ for the large grain population.

~~~~~~~~~~~~~~~~\\
{\it Stellar heating:} \hspace*{0.2cm} we assume that the disk is passively heated by stellar irradiation. The stellar 
parameters are adopted as the ``average” properties of a T Tauri star described by \citet{gullbring1998}. The stellar 
temperature and luminosity are $T_{\star}\,{=}\,4000\,\rm{K}$ and $L_{\star}\,{=}\,0.92\,L_{\odot}$ respectively. The 
stellar spectrum is modeled as a blackbody.

\begin{table}[!t]
\caption{Parameters of the reference model.}
\centering
\linespread{1.2}\selectfont
\begin{tabular}{lc}
\hline
\hline
$M_{\star}\, [M_{\odot}]$     &  0.5     \\
$T_{\star}\, [\rm K]$         &  4000    \\
$L_{\star}\, [L_{\odot}]$     &  0.92    \\
$R_{\rm in}\,[\rm{au}]$       &  0.1     \\
$R_{\rm c}\,[\rm{au}]$        &  100     \\
$R_{\rm out}\,[\rm{au}]$      &  200     \\
$\gamma$                      &  1.0     \\
$\beta$                       &  1.2     \\
$h_{\rm c}\,[\rm{au}]$        &  10      \\ 
$M_{\rm dust,tot}\, [M_{\odot}]$ & $1\,{\times}\,10^{-5}$  \\
$\Lambda$                     &  0.2     \\
$f$                           &  0.8     \\
$D\,{\rm [pc]}$               &  140     \\
\hline
\end{tabular}
\label{tab:refmodpara}
\end{table}

\subsection{Line Radiative Transfer}
\label{sec:linert}
 
As we intend to retrieve the properties of water vapor, we only simulate a mid-IR spectrum for the water lines. 
The line radiative transfer is performed with the \texttt{RADLite} code \citep{Pontoppidan2009}, which is a line 
raytracer designed to fast render spectra and images of complex line emission from protoplanetary disks. 
The molecular parameters for water are taken from the 2016 edition of the HITRAN database \citep{gordon2017}. 
The level populations are assumed in local thermodynamics equilibrium, set by the local gas temperature. 
The local line broadening is set to $0.9\,c_{s}$, where $c_{s}$ is the sound speed. 

~~~~~~~~~~~~~~~~\\
{\it Gas temperature:} \hspace*{0.2cm} we obtained the disk dust temperature by solving the problem of continuum 
radiative transfer in Sect. \ref{sec:contrt}. However, we cannot assume that the gas temperature is the same as 
the dust temperature throughout the disk as models that properly account for heating and cooling have showed that 
this assumption breaks down especially in the disk upper layers where the gas is directly exposed to ultra-violet 
and X-ray irradiation from the central star as well as the interstellar radiation field \citep[e.g.,][]{kamp2004,akimkin2013,du2014}. 
Following the approach introduced by \citet{blevins2016}, specifically their Case II, we scaled the gas temperature 
relative to the dust temperature. This approach is based on the thermo-chemical model presented in \citet{najita2011}, 
see in particular their Figure~1 which provides the scaling factor as a function of disk radius and hydrogen column 
density. The blue contours in Figure \ref{fig:gastemp} shows the resulting gas temperature of the reference model. 

~~~~~~~~~~~~~~~~\\
{\it Water vapor abundance:} \hspace*{0.2cm} for simplicity, we assumed a constant abundance for water vapor in 
the disk: $X=X_{c}(100/\rm{g2d})\,\rm{H_{2}^{-1}}$ where g2d\,=\,100 is the gas-to-dust mass ratio and $X_{c}=10^{-4}$ 
is the canonical abundance for water. The abundance is set to zero at low hydrogen column densities, $N_{\rm H}\,{\lesssim}\,10^{21}\,\rm{cm}^{-2}$, 
to account for efficient photo-destruction \citep{najita2011}. At temperatures lower than ${\sim}\,160\,\rm{K}$, the 
water is expected to condense onto dust grains. Therefore, we set the abundance at these places to a tiny value. 
The red dashed contours in Figure \ref{fig:gastemp} show the column density of water vapor measured from the disk 
upper layer to the midplane of the reference model. 

We focused on water lines in the wavelength range of $14\,{\sim}\,20\,\mu{\rm m}$, which is covered by the 
Spitzer/IRS and the upcoming JWST/MIRI instruments. The red line in Figure \ref{fig:fitrefmod} shows the 
spectrum of the reference model after convolving with a spectral resolution of 2,000, similar to that offered by JWST/MIRI.

\begin{figure}[!t]
  \centering
  \includegraphics[width=0.48\textwidth]{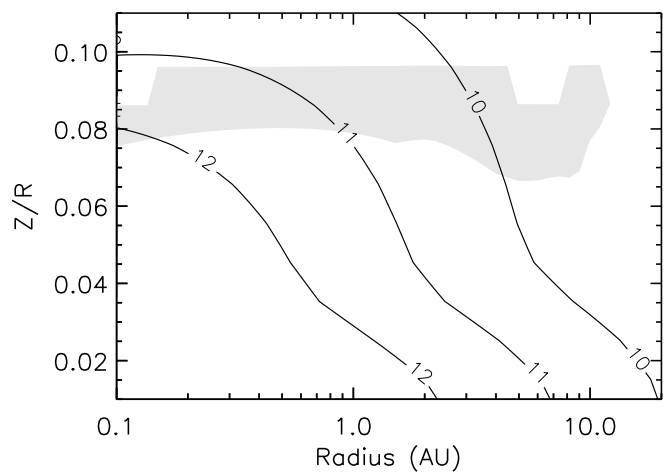}
  \caption{Gas density ${\rm Log}\,(n_{\rm gas}/{\rm cm}^{-3})$ of the reference model at 10, 11 and 12, respectively. The shaded area is
  identical to that in Figure \ref{fig:gastemp}. In the water emitting region, the gas density is close to or larger than the critical density, 
  which implies that our LTE approximation is valid.}
  \label{fig:gasdens}
\end{figure}

~~~~~~~~~~~~~~~~\\
{\it LTE approximation:} \hspace*{0.2cm} in \texttt{RADLite}, we assumed LTE excitation to simulate the reference spectrum, which makes 
the subsequent parameter retrieval fair because CLIcK calculates line emission based on the LTE assumption as well. Mid-IR 
water lines are dominated by pure rotational transitions. The critical densities for exciting these lines (at wavelengths 
considered here, $14\,{\leq}\,\lambda\,{\leq 20}\,\mu{\rm m}$) are in the range of $10^8\,{-}\,10^{11}\,{\rm cm}^{-3}$. 
In disk regions that contribute most of the water line fluxes, the gas density is comparable with the critical densities (see Figure \ref{fig:gasdens}), 
and the gas is warmer than the dust because the gas temperature is set by scaling the dust temperature based on the thermo-chemical 
model given in \citet{najita2011}. Therefore, LTE assumption is relatively reasonable in our case. It is beyond the scope of the 
paper to discuss whether or not the gas is really in LTE. Nevertheless, detailed non-LTE excitation calculations have shown that 
there is no dramatic deviation of the water line fluxes \citep[e.g.,][]{meijerink2009,woitke2018}. Similar conclusions also hold 
for mid-IR HCN and $\rm{CO_2}$ lines \citep{bruderer2015,bosman2017}. The abundances derived by LTE and non-LTE models do not differ 
by a large factor. Non-LTE excitation probably has a noticeable impact on the line shape as the emitting region can be much larger 
than in LTE models. However, observations with a high spectral resolving power \citep[e.g., $R\,{\sim}\,5000$ offered by E-ELT/METIS,][]{brandl2014} 
are required to distinguish the two scenarios.

\end{appendix}

\end{document}